\documentclass[usegraphicx,usenatbib,letterpaper, twocolappendix]{emulateapj}

\usepackage{ulem}
\usepackage{color}
\usepackage{graphics,graphicx}
\usepackage{times} 
\usepackage{amssymb}
\usepackage{amsmath}
\usepackage{natbib}
\usepackage{url}
\usepackage{multirow}
\usepackage{ctable}
\usepackage{threeparttable}
\newif\ifAMStwofonts
\AMStwofontstrue

\def\xmm{{\it XMM-Newton}}

\def\suzaku{{\it Suzaku}}

\def\swift{{\it Swift}}
\def\integral{{\it INTEGRAL~\/}}

\def\epicpn{{EPIC-pn}}
\def\epicmos1{{EPIC-MOS1}}
\def\epicmos2{{EPIC-MOS2}}
\def\epicmos{{EPIC-MOS}}

\def\nustar{{\it NuSTAR}}

\def\pcmsq{\hbox{$\rm\thinspace cm^{-2}$}}
\def\H0{{\rm ~km~s^{-1}~Mpc^{-1}}}

\def\kev{\hbox{\rm keV}}

\def\photpkevpcmsqps{\hbox{$\rm\thinspace ct~keV^{-1}~cm^{-2}~s^{-1}$}}

\def\ergpcmsqps{\hbox{$\rm\thinspace erg~cm^{-2}~s^{-1}$}}

\def\ergps{\hbox{erg~s$^{-1}$}}

\def\msun{\hbox{$\rm M_{\odot}$}}

\def\chisq{{$\chi^{2}$}}

\def\xspec{\hbox{\small XSPEC}}
\def\xspecv{\hbox{\small XSPEC}\, v12.6.0f}

\def\heasoft{\hbox{\rm{\small HEASOFT}}}
\def\nustardas{\rm {\small NUSTARDAS}}

\def\grid25{\hbox{\rm{\small GRID25}}}

\def\simpl{\rm{\small SIMPL}}

\def\tbabs{\rm{\small TBABS}}

\def\diskbb{\rm{\small DISKBB}}
\def\diskpbb{\rm{\small DISKPBB}}

\def\comptt{\rm{\small COMPTT}}

\def\cutoffpl{\rm{\small CUTOFFPL}}

\def\fexxv{\hbox{\rm Fe\,{\small XXV}}}
\def\fexxvi{\hbox{\rm Fe\,{\small XXVI}}}
\def\fei{\hbox{\rm Fe\,{\small I}}}

\def\ka{K\,$\alpha$}

\def\eg{{\it e.g.}}

\def\ie{{\it i.e.~\/}}

\def\la{\mathrel{\hbox{\rlap{\hbox{\lower4pt\hbox{$\sim$}}}{\raise2pt\hbox{$<$}}}}}
\def\ga{\mathrel{\hbox{\rlap{\hbox{\lower4pt\hbox{$\sim$}}}{\raise2pt\hbox{$>$}}}}}

\def\d25{D$_{25}$}
\def\nh{{$N_{\rm H}$}}

\def\.25{0.25 keV\thinspace}

\def\rg{$R_{\rm G}$}

\def\hoix{\rm Holmberg\,IX X-1}
\def\elow{6.0}
\def\ehigh{10.0}

\shorttitle{Broadband X-ray observations of \hoix}
\shortauthors{D.~J. Walton et al.}

\begin{document}

\title{The Broadband Spectral Variability of Holmberg\,IX X-1}

\author{D. J. Walton\altaffilmark{1,2,3},
F. F\"urst\altaffilmark{2,4},
F. A. Harrison\altaffilmark{2},
M. J. Middleton\altaffilmark{5},
A. C. Fabian\altaffilmark{3},
M. Bachetti\altaffilmark{6},
D. Barret\altaffilmark{7,8}, \\
J. M. Miller\altaffilmark{9},
A. Ptak\altaffilmark{10},
V. Rana\altaffilmark{2},
D. Stern\altaffilmark{1},
L. Tao\altaffilmark{2}
}
\affil{
$^{1}$ Jet Propulsion Laboratory, California Institute of Technology, Pasadena, CA 91109, USA \\
$^{2}$ Space Radiation Laboratory, California Institute of Technology, Pasadena, CA 91125, USA \\
$^{3}$ Institute of Astronomy, University of Cambridge, Madingley Road, Cambridge CB3 0HA, UK \\
$^{4}$ European Space Astronomy Centre (ESA/ESAC), Science Operations Department, Villanueva de la Cañada (Madrid), Spain \\
$^{5}$ Department of Physics and Astronomy, University of Southampton, Highfield, Southampton SO17 1BJ, UK \\
$^{6}$ INAF/Osservatorio Astronomico di Cagliari, via della Scienza 5, I-09047 Selargius (CA), Italy \\
$^{7}$ Universite de Toulouse; UPS-OMP; IRAP; Toulouse, France \\
$^{8}$ CNRS; IRAP; 9 Av. colonel Roche, BP 44346, F-31028 Toulouse cedex 4, France \\
$^{9}$ Department of Astronomy, University of Michigan, 1085 S. University Ave., Ann Arbor, MI, 49109-1107, USA \\
$^{10}$ NASA Goddard Space Flight Center, Greenbelt, MD 20771, USA \\
}

\begin{abstract}
We present results from four new broadband X-ray observations of the extreme
ultraluminous X-ray source \hoix\ ($L_{\rm{X}} > 10^{40}$\,\ergps), performed by
\suzaku\ and \nustar\ in coordination. Combined with the archival data, we now have
broadband observations of this remarkable source from six separate epochs. Two of
these new observations probe lower fluxes than seen previously, allowing us to
extend our knowledge of the broadband spectral variability exhibited. The spectra are
well fit by two thermal blackbody components, which dominate the emission below
10\,keV, as well as a steep ($\Gamma \sim 3.5$) powerlaw tail which dominates
above $\sim$15\,keV. Remarkably, while the 0.3--10.0\,keV flux varies by a factor of
$\sim$3 between all these epochs, the 15--40\,keV flux varies by only $\sim$20\%.
Although the spectral variability is strongest in the $\sim$1--10\,keV band, both of the
thermal components are required to vary when all epochs are considered. We also
re-visit the search for iron absorption features, leveraging the high-energy \nustar\
data to improve our sensitivity to extreme velocity outflows in light of the ultra-fast
outflow recently detected in NGC\,1313 X-1. Iron absorption from a similar outflow
along our line of sight can be ruled out in this case. We discuss these results in the
context of super-Eddington accretion models that invoke a funnel-like geometry for
the inner flow, and propose a scenario in which we have an almost face-on view of a
funnel that expands to larger radii with increasing flux, resulting in an increasing
degree of geometrical collimation for the emission from intermediate temperature
regions.
\end{abstract}

\begin{keywords}
{Black hole physics -- X-rays: binaries -- X-rays: individual (\hoix)}
\end{keywords}

\section{Introduction}

\hoix\ is one of the best studied members of the ultraluminous X-ray source (ULX)
population (see \citealt{Feng11rev} for a recent review on ULXs), as it is one of the few 
sources within $\sim$5\,Mpc to persistently radiate at an extreme X-ray luminosity of
$L_{\rm X} > 10^{40}$\,\ergps\ (\eg\ \citealt{Kong10, Vierdayanti10, WaltonULXCat};
the distance to Holmberg\,IX is 3.55\,Mpc, \citealt{Paturel02}). Early observations of
\hoix\ with \xmm\ (\citealt{XMM}) suggested the presence of a cool accretion disk
($T_{\rm{in}} \sim 0.3$\,keV; \citealt{Miller03, Miller13ulx}), and a hard powerlaw
continuum at higher energies ($>$2\,keV), consistent with the expectation for
low-Eddington rate accretion. This suggested the presence of a massive
($>$100\,\msun) black hole accreting from a standard thin disk. However, higher
quality data subsequently revealed deviations from a powerlaw continuum in the
$\sim$2--10\,keV band (\eg\ \citealt{Stobbart06, Gladstone09, Walton13hoIXfeK}),
calling this identification into question. 

Most recently, our coordinated broadband observations  of \hoix\ with \xmm, \suzaku\
(\citealt{SUZAKU}) and \nustar\ (\citealt{NUSTAR}) robustly confirmed that the X-ray
continuum above $\sim$2\,keV does not have a powerlaw form, in good agreement
with the weak hard X-ray detection revealed by \integral\ (\citealt{Sazonov14}).
Instead, the hard X-ray spectrum appears to be dominated by a second, hotter ($T
\sim 3$\,keV, at least at lower fluxes) multi-color blackbody emission component likely
associated with the inner accretion disk, similar to the rest of the ULX population
observed by \nustar\ to date (\eg\ \citealt{Bachetti13, Walton13culx, Walton15,
Walton15hoII, Rana15, Mukherjee15}), with a steep powerlaw tail also potentially
detected at the highest energies (\citealt{Walton14hoIX}). The broadband spectra
observed are generally consistent with the expectation for super-Eddington accretion
onto a compact stellar remnant (\eg\ \citealt{Shakura73, Dotan11}). Indeed, we now
know that at least three of this population of extreme ULXs are actually highly
super-Eddington neutron stars (\citealt{Bachetti14nat, Fuerst16p13, Israel16,
Israel17p13}). Furthermore, in the case of NGC\,1313 X-1, another $L_{\rm X} \sim
10^{40}$\,\ergps\ ULX, discrete absorption features from a massive outflow have now
finally been detected (\citealt{Pinto16nat, Walton16ufo}), following an initial suggestion
that low-energy ($\sim$1\,keV) residuals could be related to such a wind
(\citealt{Middleton14, Middleton15soft}). Large outflows from the accretion flow are
broadly expected from super-Eddington accretion theory (\eg\ \citealt{Poutanen07}).

The most remarkable aspect of these broadband observations, however, is the
unusual spectral variability. In our initial program, we obtained data from two epochs
separated by $\sim$2 weeks. During the second epoch, \hoix\ was significantly brighter
than during the first by a factor of $\sim$2. This was primarily driven by an increase in
the $\sim$1--15\,keV bandpass, while the fluxes below $\sim$1\,keV and above
$\sim$15\,keV remained relatively constant, resulting in a significantly more peaked
spectrum in the second observation (\citealt{Walton14hoIX}). The similar fluxes at the
highest and lowest energies of the observed 0.3--40\,keV bandpass implies that the
variability is dominated by one of the two multi-color blackbody components, but with
only two epochs it was not clear which; in either case, the observed evolution would be
highly non-standard behavior for blackbody emission.

In order to further investigate the spectral variability exhibited by \hoix, we obtained
four additional epochs of coordinated broadband observations with \suzaku\ and
\nustar. Here, we present the results obtained with these new observations (hereafter
epochs 3--6), along with a re-analysis of the archival data (epochs 1--2). This paper
is structured as follows: section \ref{sec_red} describes our data reduction procedure,
and section \ref{sec_spec} describes our spectral analysis of the existing broadband
datasets (both new and archival). Finally, we discuss our results and present our
conclusions in section \ref{sec_dis}.

\section{Observations and Data Reduction}
\label{sec_red}

Table \ref{tab_obs} provides the details of the X-ray observations of \hoix\ presented in
this work. Our data reduction largely follows the procedures outlined in
\cite{Walton14hoIX}. The \xmm\ data for epochs 1 and 2 were re-reduced using SAS
v14.0.0 in exactly the same manner as described in our previous work for both the
\epicpn\ and \epicmos\ detectors (\citealt{XMM_PN, XMM_MOS}), using more
recent calibration files (up-to-date as of July 2015). Our new \suzaku\ observations
(epochs 3--6) were also reduced with \heasoft\ v6.18 in the same manner as the epoch
1 data; as before we only utilize the XIS detectors (\citealt{SUZAKU_XIS}) given the
high-energy \nustar\ coverage. The only major update to our data reduction procedure
in comparison to \cite{Walton14hoIX} is for the \nustar\ data. The standard science
data (mode 1) were reduced with \nustardas\ v1.5.1 in the same manner as in that
work, again with updated calibration files (\nustar\ CALDB 20150316). However, in
order to maximize the S/N, in this work we also make use of the spacecraft science
data (mode 6) following the procedure outlined in \citet[][see also the
\nustar\ Users
Guide\footnote{http://heasarc.gsfc.nasa.gov/docs/nustar/analysis/nustar\_swguide.pdf}]{Walton16cyg}.
In brief, spacecraft science mode refers to data recorded during periods of the
observation in which the X-ray source is still visible, but the star tracker on the optics
bench cannot return a good aspect solution. During such times, the star trackers on
the spacecraft bus are utilized instead. For these observations, this provides an
additional $\sim$20--40\% good exposure (depending on the observation). The
inclusion of the mode 6 data allows us to fit the \nustar\ data for \hoix\ from 3 to
40\,keV for all epochs. As in \cite{Walton14hoIX}, we model the \xmm\ spectra over
the 0.3--10.0\,keV range, and the \suzaku\ data over the 0.6--10.0\,keV range for the
front-illuminated XIS units (XIS0, XIS3) and the 0.7--9.0\,keV range for the
back-illuminated XIS1, excluding 1.7--2.1\,keV owing to known calibration issues
around the instrumental edges.

\begin{table}
  \caption{Details of the X-ray observations of Holmberg\,IX X-1 considered in this
  work, ordered chronologically.}
\begin{center}
\begin{tabular}{c c c c}
\hline
\hline
\\[-0.1cm]
Mission & OBSID & Date & Good Exposure\tmark[a] \\
& & & (ks) \\
\\[-0.2cm]
\hline
\hline
\\[-0.1cm]
\multicolumn{4}{c}{\textit{Epoch 1 (medium flux)}} \\
\\[-0.2cm]
\suzaku\ & 707019020 & 2012-10-21 & 107 \\
\\[-0.225cm]
\xmm\ & 0693850801 & 2012-10-23 & 6/10 \\
\\[-0.225cm]
\suzaku\ & 707019030 & 2012-10-24 & 107 \\
\\[-0.225cm]
\xmm\ & 0693850901 & 2012-10-25 & 7/13 \\
\\[-0.225cm]
\suzaku\ & 707019040 & 2012-10-26 & 110 \\
\\[-0.225cm]
\nustar\ & 30002033002 & 2012-10-26 & 43 (12) \\
\\[-0.225cm]
\nustar\ & 30002033003 & 2012-10-26 & 124 (36) \\
\\[-0.225cm]
\xmm\ & 0693851001 & 2012-10-27 & 4/13 \\
\\[-0.1cm]
\multicolumn{4}{c}{\textit{Epoch 2 (high flux)}} \\
\\[-0.2cm]
\nustar\ & 30002033005 & 2012-11-11 & 49 (8) \\
\\[-0.225cm]
\nustar\ & 30002033006 & 2012-11-11 & 41 (6) \\
\\[-0.225cm]
\xmm\ & 0693851701 & 2012-11-12 & 7/9 \\
\\[-0.225cm]
\nustar\ & 30002033008 & 2012-11-14 & 18 (3) \\
\\[-0.225cm]
\xmm\ & 0693851801 & 2012-11-14 & 7/9 \\
\\[-0.225cm]
\nustar\ & 30002033010 & 2012-11-15 & 59 (10) \\
\\[-0.225cm]
\xmm\ & 0693851101 & 2012-11-16 & 3/7 \\
\\[-0.1cm]
\multicolumn{4}{c}{\textit{Epoch 3 (medium flux)}} \\
\\[-0.2cm]
\nustar\ & 30002034002 & 2014-05-02 & 82 (24) \\
\\[-0.225cm]
\suzaku\ & 707019010 & 2014-05-03 & 32 \\
\\[-0.1cm]
\multicolumn{4}{c}{\textit{Epoch 4 (low flux)}} \\
\\[-0.2cm]
\nustar\ & 30002034004 & 2014-11-15 & 81 (20) \\
\\[-0.225cm]
\suzaku\ & 707019020 & 2014-11-15 & 34 \\
\\[-0.1cm]
\multicolumn{4}{c}{\textit{Epoch 5 (low flux)}} \\
\\[-0.2cm]
\nustar\ & 30002034006 & 2015-04-06 & 64 (14) \\
\\[-0.225cm]
\suzaku\ & 707019030 & 2015-04-06 & 32 \\
\\[-0.1cm]
\multicolumn{4}{c}{\textit{Epoch 6 (medium flux)}} \\
\\[-0.2cm]
\nustar\ & 30002034008 & 2015-05-16 & 67 (10) \\
\\[-0.225cm]
\suzaku\ & 707019040 & 2015-05-16 & 34 \\
\\[-0.2cm]
\hline
\hline
\\[-0.15cm]
\end{tabular}
\\
$^{a}$ \xmm\ exposures are listed for the \epicpn/MOS detectors, while the
\nustar\ exposures quoted are the total for each of the focal plane modules,
with the mode 6 contribution in parentheses, and the \suzaku\ exposures
quoted are for each of the operational XIS units.
\vspace*{0.3cm}
\label{tab_obs}
\end{center}
\end{table}

In the following sections, we perform spectral analysis with \xspecv\ (\citealt{xspec}),
and quote parameter uncertainties at the 90\% confidence level for one interesting
parameter. Cross-calibration uncertainties between the different detectors are
accounted for by allowing multiplicative constants to float between the datasets from
a given epoch, fixing FPMA at unity. The values are within $\sim$10\% (or less) of
unity, as expected (\citealt{NUSTARcal}).

\begin{figure*}
\hspace*{-0.5cm}
\epsscale{0.55}
\plotone{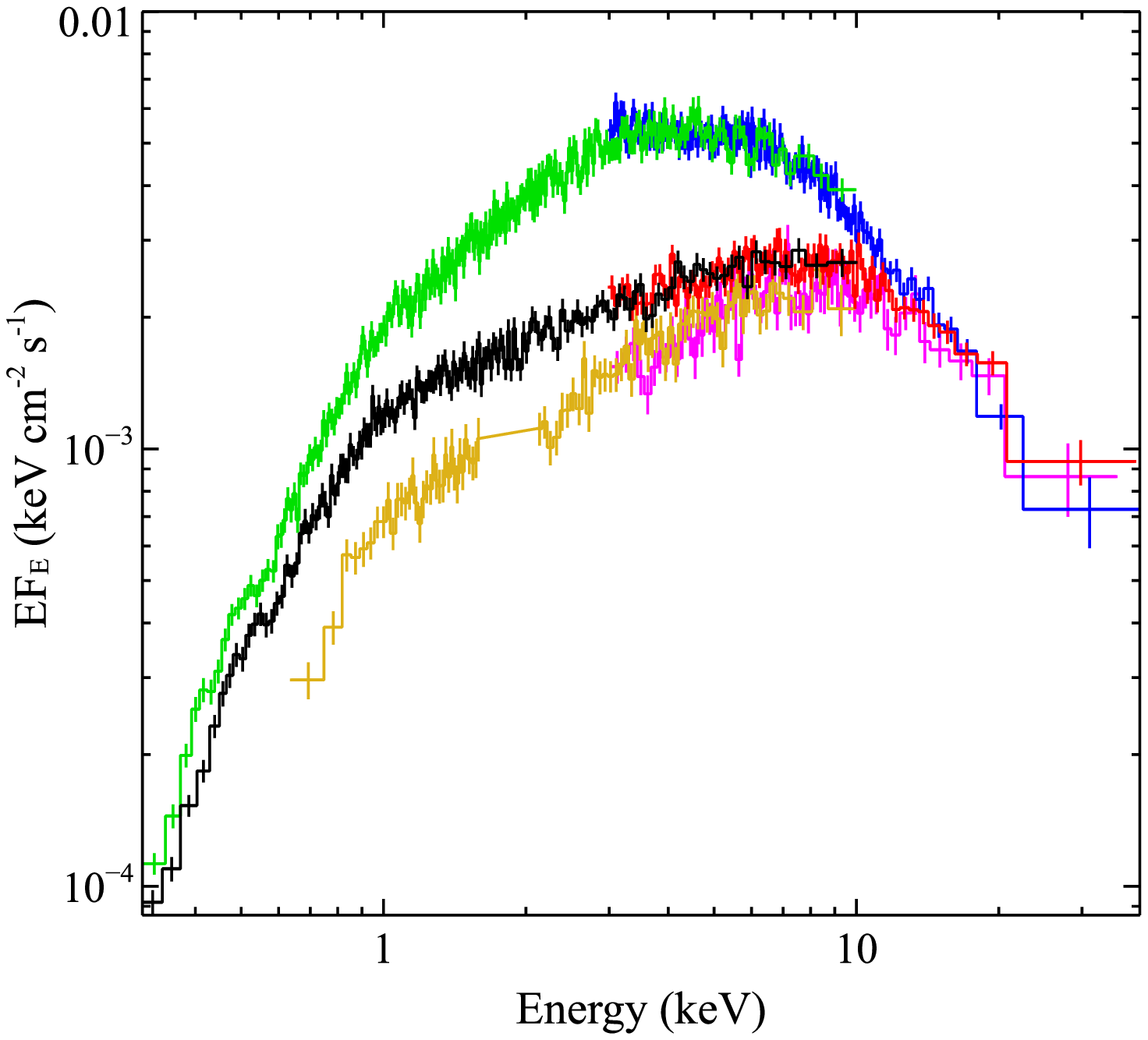}
\hspace{0.5cm}
\plotone{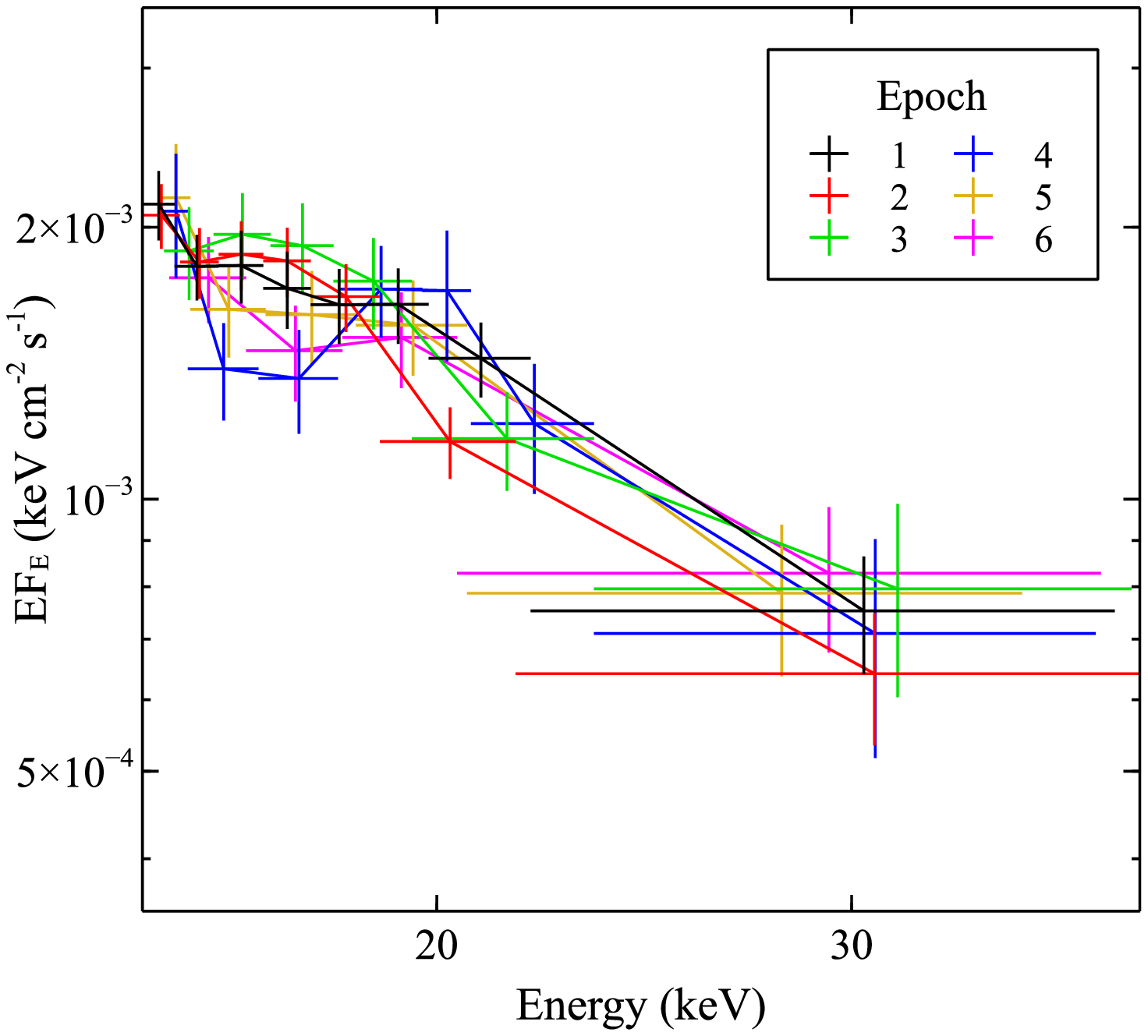}
\caption{\textit{Left:} The broadband spectral evolution displayed by \hoix. We
show the data from epochs 1, 2 and 4, being representative of the three `states'
(medium, high and low, respectively) now covered with broadband X-ray
observations. For clarity, we only show the \nustar\ data from FPMA (red, blue and
magenta for epochs 1, 2 and 4, respectively) along with one of the accompanying
soft X-ray detectors (black, green and orange for epochs 1, 2 and 4, respectively,
with epochs 1 and 2 showing data from \epicpn\ on \xmm, and epoch 4 showing
the FI XIS units on \suzaku). Of the three other new broadband epochs presented
in this work, two are similar to epoch 1 (epochs 3 and 6) and the other (epoch 5) is
similar to epoch 4. \textit{Right:} The high-energy ($>$15\,keV) data from \nustar\
(again, FPMA only for clarity). Here we show the data from all six epochs,
demonstrating the relative lack of variability observed above $\sim$15\,keV in
comparison to the lower energy data. All the data have been unfolded through the
same model, which is simply constant with energy, and the data have been further
rebinned for visual purposes.}
\vspace{0.3cm}
\label{fig_spec}
\end{figure*}

\section{Spectral Analysis}
\label{sec_spec}

The primary goal of our follow-up \suzaku+\nustar\ observations was to obtain
broadband observations of \hoix\ in flux states that differ from those presented in
\cite{Walton14hoIX}, such that we can improve our understanding of the unusual
broadband spectral variability exhibited by this source. Of the additional four epochs
we obtained, the first (epoch 3) was found to be extremely similar to the fainter of
the two states observed in our original observations (epoch 1). The following two
(epochs 4 and 5) were also very similar to each other, but distinct from epochs 1--3,
probing even lower fluxes. Finally, the last (epoch 6) was again broadly similar to
epochs 1 and 3, although epoch 6 is slightly fainter overall and so is not completely
identical.

Throughout this work, we group the observations into low (epochs 4 and 5), medium
(epochs 1, 3 and 6) and high (epoch 2) states, as indicated in Table \ref{tab_obs}.
The broadband spectrum from epoch 4 is shown in comparison to epochs 1 and 2 in
Figure \ref{fig_spec} (\textit{left panel}). The evolution from the flux exhibited in epoch
1 to the even lower fluxes observed during epochs 4 and 5 is primarily dominated by
variability at low energies. Remarkably, the high-energy data appears to remain
extremely similar for all six epochs probed by \nustar\ to date, as also shown in Figure
\ref{fig_spec} (\textit{right panel}). Indeed, if we fit the 15--40~keV \nustar\ data for all
the epochs with a simple powerlaw model, we find that the photon indices are all
consistent within their $\sim$90\% confidence limits ($\Gamma\sim3.5$), and the
15--40~keV fluxes only vary by $\sim$20\%, despite the factor of $>$2 differences
seen below 10\,keV (\citealt{Walton14hoIX}).

\subsection{Broadband Spectral Variability}

In order to characterize the spectra and study the observed spectral variability, we fit
each of the six epochs with a common model, such that the results from each can be
directly compared. In order to obtain the tightest constraints, and determine which of
the model parameters are important in producing the spectral variability, we analyse
the broadband data from all six epochs simultaneously.

We focus on the model presented in \cite{Walton14hoIX} that combines \diskbb,
\diskpbb\ and \simpl\ (\citealt{diskbb, diskpbb, simpl}), which was able to successfully
describe the first two broadband epochs. We refer the reader to \cite{Walton14hoIX}
for full details, but in brief the \diskbb\ and \diskpbb\ components account for the two
thermal components that dominate the spectrum below $\sim$15\,keV (the former
assumes the thin disk model outlined in \citealt{Shakura73}, while the latter allows
for an accretion disk continuum with a variable radial temperature index), with \simpl\
applied to the higher-temperature \diskpbb\ component to provide the high-energy
powerlaw tail that is present above $\sim$15\,keV. In addition to these continuum
components, the model also includes two neutral absorption components, modeled
with \tbabs. We use the abundance set of \cite{tbabs} and the cross-sections of
\cite{Verner96}. The first of these components is fixed at the Galactic column
($N_{\rm{H,Gal}} = 5.54\times10^{20}$ cm$^{-2}$; \citealt{NH}), and the second is
assumed to be intrinsic to either the Holmberg\,IX galaxy ($z=0.000153$) or to the
\hoix\ system itself.


\begin{table*}
  \caption{Results obtained for the dual-thermal model combining the DISKBB and DISKPBB
  multicolor blackbody disk compoents for the six broadband epochs currently available for
  Holmberg\,IX X-1}
  \vspace{-0.25cm}
\begin{center}
  \hspace*{-0.1cm}
\begin{tabular}{c c c c c c c c c c}
\hline
\hline
\\[-0.1cm]
Model & \multicolumn{2}{c}{Parameter} & Global & \multicolumn{6}{c}{Epoch} \\
\\[-0.2cm]
Component & & & & 1 & 2 & 3 & 4 & 5 & 6 \\
\\[-0.2cm]
\hline
\hline
\\[-0.1cm]
\multicolumn{10}{c}{Constant \diskbb\ temperature} \\
\\[-0.1cm]
\tbabs\ & $N_{\rm{H,int}}$ & [$10^{21}$ cm$^{-2}$] & $1.45 \pm 0.08$ \\
\\[-0.2cm]
\diskbb & $T_{\rm{in}}$ & [keV] & $0.29 \pm 0.02$ &  &  \\
\\[-0.2cm]
& Norm & & & $11^{+3}_{-2}$ & $14^{+6}_{-4}$ & $17^{+6}_{-4}$ & $11^{+4}_{-3}$ & $11 \pm 3$ & $4^{+3}_{-2}$ \\
\\[-0.2cm]
\diskpbb\ & $T_{\rm{in}}$ & [keV] & & $2.6^{+0.4}_{-0.1}$ & $1.3^{+0.9}_{-0.1}$ & $2.6^{+0.3}_{-0.2}$ & $2.6^{+0.4}_{-0.1}$ & $2.6^{+0.3}_{-0.2}$ & $2.9^{+0.3}_{-0.2}$ \\
\\[-0.2cm]
& $p$ & & & $0.574^{+0.005}_{-0.004}$ & $0.69^{+0.03}_{-0.02}$ & $0.60 \pm 0.01$ & $0.63^{+0.01}_{-0.02}$ & $0.64^{+0.02}_{-0.01}$ & $0.56 \pm 0.01$ \\
\\[-0.2cm]
& Norm & [$10^{-3}$] & & $3.2^{+0.6}_{-1.1}$ & $190^{+70}_{-80}$ & $3.8^{+1.4}_{-1.5}$ & $3.5^{+1.2}_{-1.4}$ & $4.2^{+1.6}_{-1.7}$ & $1.8 \pm 0.7$ \\
\\[-0.2cm]
\simpl\ & $\Gamma$ & & $3.68^{+0.07}_{-0.05}$ \\
\\[-0.2cm]
& $f_{\rm{scat}}$ & [\%] & $>67$ \\
\\[-0.2cm]
\hline
\\[-0.1cm]
\chisq/DoF & & & 8792/8515 \\
\\[-0.2cm]
\hline
\hline
\\[-0.1cm]
\multicolumn{10}{c}{Constant \diskpbb\ temperature} \\
\\[-0.1cm]
\tbabs\ & $N_{\rm{H,int}}$ & [$10^{21}$ cm$^{-2}$] & $1.50 \pm 0.07$ & \\
\\[-0.2cm]
\diskbb & $T_{\rm{in}}$ & [keV] & & $0.28^{+0.02}_{-0.01}$ & $1.70^{+0.04}_{-0.03}$ & $0.26 \pm 0.02$ & $0.30 \pm 0.03$ & $0.29 \pm 0.03$ & $0.41 \pm 0.06$ \\
\\[-0.2cm]
& Norm & & & $12^{+4}_{-3}$ & $0.063 \pm 0.005$ & $27^{+15}_{-9}$ & $10^{+5}_{-3}$ & $10^{+6}_{-4}$ & $2.2^{+1.8}_{-0.9}$ \\
\\[-0.2cm]
\diskpbb\ & $T_{\rm{in}}$ & [keV] & $2.9^{+0.4}_{-1.1}$ \\
\\[-0.2cm]
& $p$ & & & $0.571 \pm 0.004$ & $0.542 \pm 0.005$ & $0.59 \pm 0.01$ & $0.63^{+0.01}_{-0.02}$ & $0.63 \pm 0.01$ & $0.58 \pm 0.01$ \\
\\[-0.2cm]
& Norm & [$10^{-3}$] & & $2.3 \pm 0.9$ & $1.7^{+0.6}_{-0.7}$ & $2.5^{+1.1}_{-0.8}$ & $2.6^{+1.3}_{-0.9}$ & $2.7^{+1.3}_{-1.0}$ & $2.1^{+1.0}_{-0.9}$ \\
\\[-0.2cm]
\simpl\ & $\Gamma$ & & $>3.4$\tmark[a] \\
\\[-0.2cm]
& $f_{\rm{scat}}$ & [\%] & $>41$ \\
\\[-0.2cm]
\hline
\\[-0.1cm]
\chisq/DoF & & & 8789/8515 & \\
\\[-0.2cm]
\hline
\hline
\\[-0.1cm]
$F_{0.3-40.0}$\tmark[b] & \multicolumn{2}{c}{\multirow{5}{*}{[$10^{-12}$\,\ergpcmsqps]}} & & $12.8^{+0.2}_{-0.1}$ & $22.3 \pm 0.2$ & $12.4 \pm 0.2$ & $10.1 \pm 0.2$ & $9.8 \pm 0.2$ & $11.5 \pm 0.2$ \\
\\[-0.2cm]
$F_{0.3-1.0}$\tmark[b] & & & & $1.02 \pm 0.02$ & $1.47 \pm 0.02$ & $1.01 \pm 0.03$ & $0.66 \pm 0.03$ & $0.60 \pm 0.03$ & $0.87 \pm 0.03$ \\
\\[-0.2cm]
$F_{1.0-10.0}$\tmark[b] & & & & $8.6 \pm 0.1$ & $17.5 \pm 0.2$ & $8.2 \pm 0.2$ & $6.6 \pm 0.1$ & $6.3 \pm 0.1$ & $7.8 \pm 0.2$ \\
\\[-0.2cm]
$F_{10.0-40.0}$\tmark[b] & & & & $3.23 \pm 0.08$ & $3.17 \pm 0.09$ & $3.29 \pm 0.12$ & $3.01 \pm 0.12$ & $2.91 \pm 0.12$ & $3.07 \pm 0.13$ \\
\\[-0.2cm]
\hline
\\[-0.1cm]
$L_{0.3-40.0}$\tmark[c] & \multicolumn{2}{c}{[$10^{40}$\,\ergps]} & & $2.30 \pm 0.03$ & $3.84 \pm 0.05$ & $2.23 \pm 0.05$ & $1.77 \pm 0.04$ & $1.68 \pm 0.04$ & $2.08 \pm 0.05$ \\
\\[-0.2cm]
\hline
\hline
\end{tabular}
\vspace{-0.2cm}
\label{tab_param}
\end{center}
$^{a}$ We set an upper limit for the photon index of $\Gamma \leq 4.0$. \\
$^{b}$ The observed flux in the full 0.3--40.0\,keV band, and the 0.3--2.0, 2.0--10.0 and 10.0--40.0\,keV sub-bands, respectively (consistent for both models). \\
$^{c}$ De-absorbed luminosity in the full 0.3--40.0\,keV band. These values assume isotropic emission, and may therefore represent upper limits (see Section \ref{sec_dis}).
\vspace{0.2cm}
\end{table*}

\begin{figure*}
\hspace*{-0.5cm}
\epsscale{0.55}
\plotone{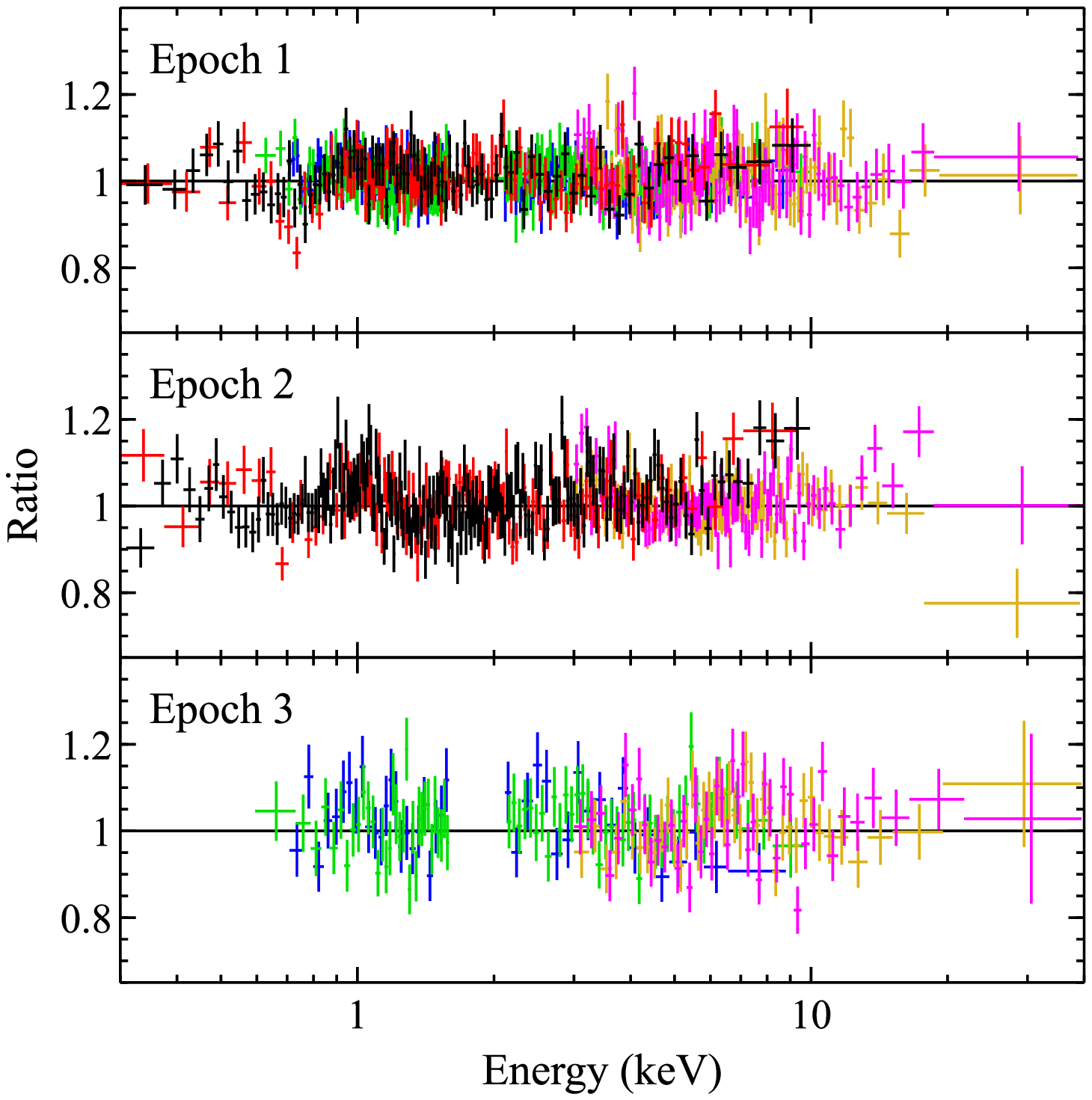}
\hspace{0.5cm}
\plotone{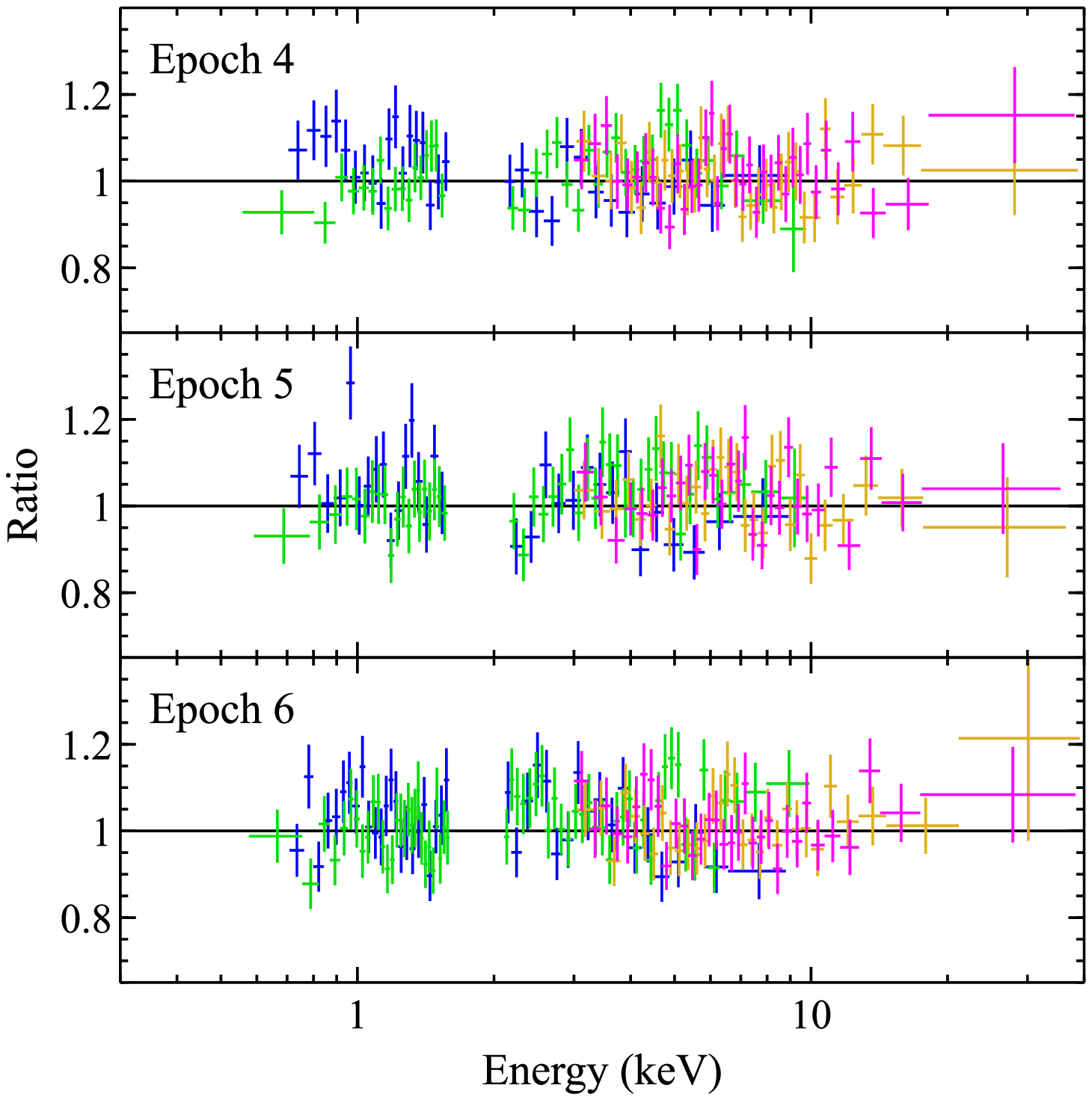}
\caption{Data/model ratios for the constant DISKPBB temperature model
presented in Table \ref{tab_param}. \epicpn, \epicmos, FI XIS, BI XIS, FPMA
and FPMB data are shown in black, red, green, blue, magenta and orange,
respectively, and the data have been further rebinned for visual purposes.}
\vspace{0.3cm}
\label{fig_specrat}
\end{figure*}

As a starting point, we apply our model with the high-energy photon index of the
\simpl\ component and the neutral absorption column intrinsic to Holmberg\,IX
linked between the different epochs (but still free to vary globally). We link the
former following our analysis of the 15--40\,keV spectrum above, while we link the
latter because for the two broadband epochs with \xmm\ coverage down to 0.3\,keV
(epochs 1 and 2) we found the neutral absorption column to be consistent in our
prior work on these data (see also \citealt{Miller13ulx}, who find that a constant
column provides a good fit to all the archival \xmm\ observations of \hoix), and the
remaining epochs only have coverage down to 0.7\,keV, reducing their sensitivity to
small variations in \nh. Following \cite{Walton14hoIX}, we restrict the range over
which the photon index can vary to be $1.5 \leq \Gamma \leq 4.0$. The parameters
that are initially allowed to vary from epoch to epoch are therefore the inner
temperatures ($T_{\rm{in}}$) and normalizations of the \diskbb\ and \diskpbb\
components, the radial temperature index ($p$) of the \diskpbb\ component, and
the scattered fraction ($f_{\rm{scat}}$) for the \simpl\ component, which acts as an
effective normalization for the high-energy powerlaw tail. This model provides a good
global fit to the six available broadband epochs: $\chi^{2}$ =  8779 for 8505 degrees
of freedom (DoF).

From this baseline, we proceed to investigate which of the other key model
parameters drive the observed variability by trying to link them across each of the
epochs (but again allowing them to vary globally), beginning with the temperatures
of each of the \diskbb\ and \diskpbb\ components in turn. Neither of these scenarios
significantly degrades the quality of the fit, and we find that both provide very similar
fits, with a constant temperature for the hotter \diskpbb\ component marginally
preferred: fixing the \diskbb\ temperature results in a fit of \chisq/DoF = 8792/8510
(worse by $\Delta\chi^{2}=13$ for 5 fewer free parameters), and fixing the \diskpbb\
temperature results in a fit of \chisq/DoF = 8786/8510 (worse by $\Delta\chi^{2}=7$,
also for 5 fewer free parameters). However, if we try to fix \textit{both} of the
temperatures across all epochs, the global fit does worsen significantly, resulting in
\chisq/DoF = 8931/8515. This is because, as discussed in \cite{Walton14hoIX}, with
this simple parameterization the temperature of one of the two thermal components
must evolve to explain the observed differences between epochs 1 and 2.

Next we consider the radial temperature index for the \diskpbb\ component.
Regardless of which of the \diskbb\ or \diskpbb\ temperatures we choose to link,
additionally linking $p$ significantly worsens the fit. With a constant temperature
for the cooler \diskbb\ component, the fit degrades to \chisq/DoF = 8952/8515
($\Delta\chi^{2}$ = 160, 5 fewer free parameters), and with a constant
temperature for the hotter \diskpbb\ component the fit degrades to \chisq/DoF =
8890/8515 ($\Delta\chi^{2}$ = 104). This is also the case if we do not link either
of the temperatures at all, in which case the best fit is \chisq/DoF = 8850/8510
($\Delta\chi^{2}$ = 71, compared to the fit with both temperatures free), so the
observed variation in $p$ is not merely a consequence of this choice. The radial
temperature index $p$ determines the slope of the \diskpbb\ model up to the
point at which it falls away with a Wien spectrum (set by $T_{\rm{in}}$). This is
critical in determining the slope of the observed spectrum in the $\sim$1--7\,keV
energy range for the lower flux states, given the low temperatures
($\sim$0.3\,keV) obtained for the \diskbb\ component. Therefore, even if one
tries to assign the majority of the variability to the lower temperature \diskbb\
component by requiring that the \diskpbb\ temperature is constant, the spectral
form of the this hotter component is still required to vary. While the values vary,
we consistently find that $p < 0.75$ (the value expected for a thin disk;
\citealt{Shakura73}), which would suggest a disk in which photon advection is
important (\eg\ \citealt{Abram88}).

\begin{figure}
\hspace*{-0.7cm}
\epsscale{1.15}
\plotone{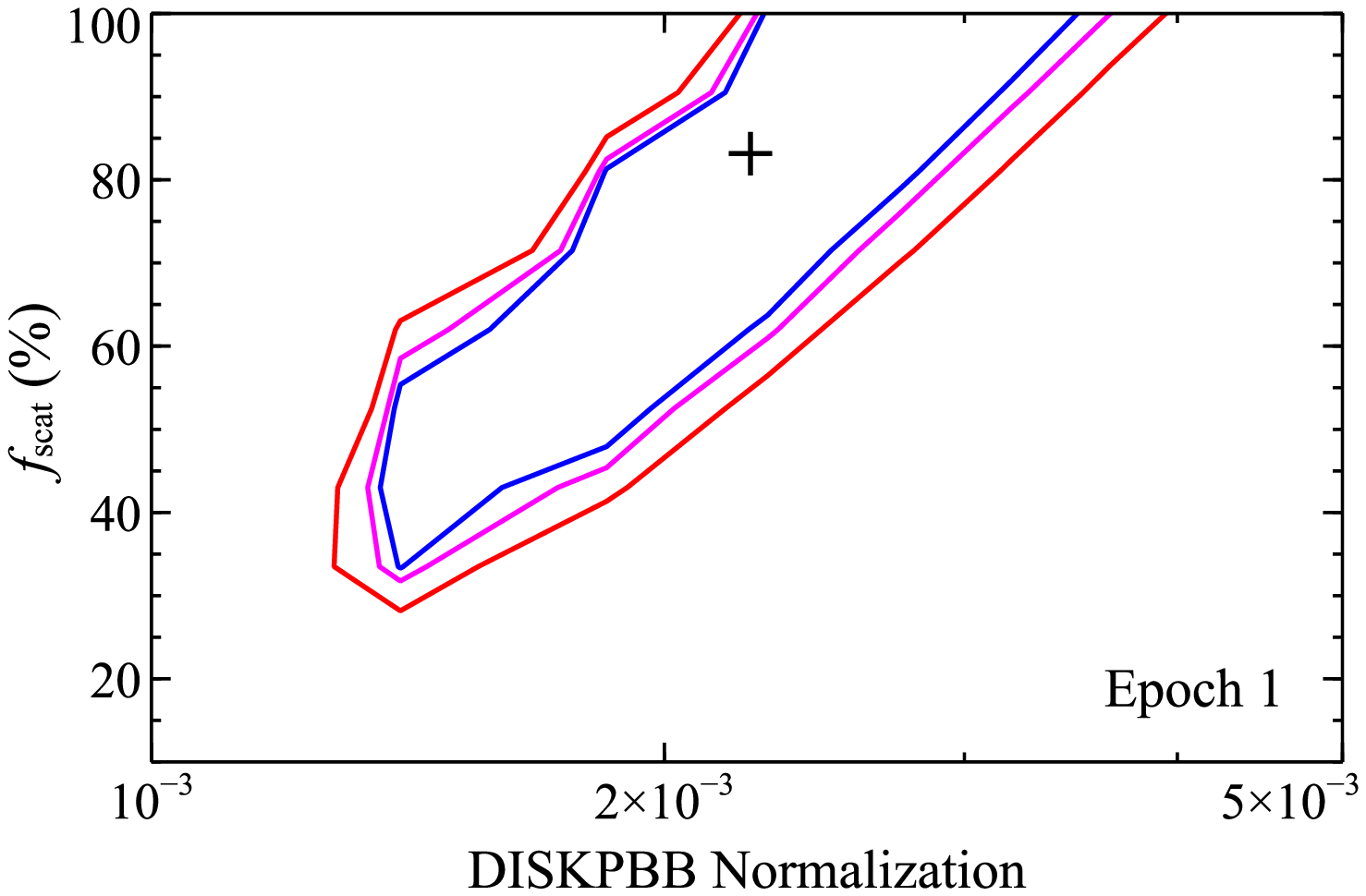}
\hspace*{-0.7cm}
\vspace{0.25cm}
\plotone{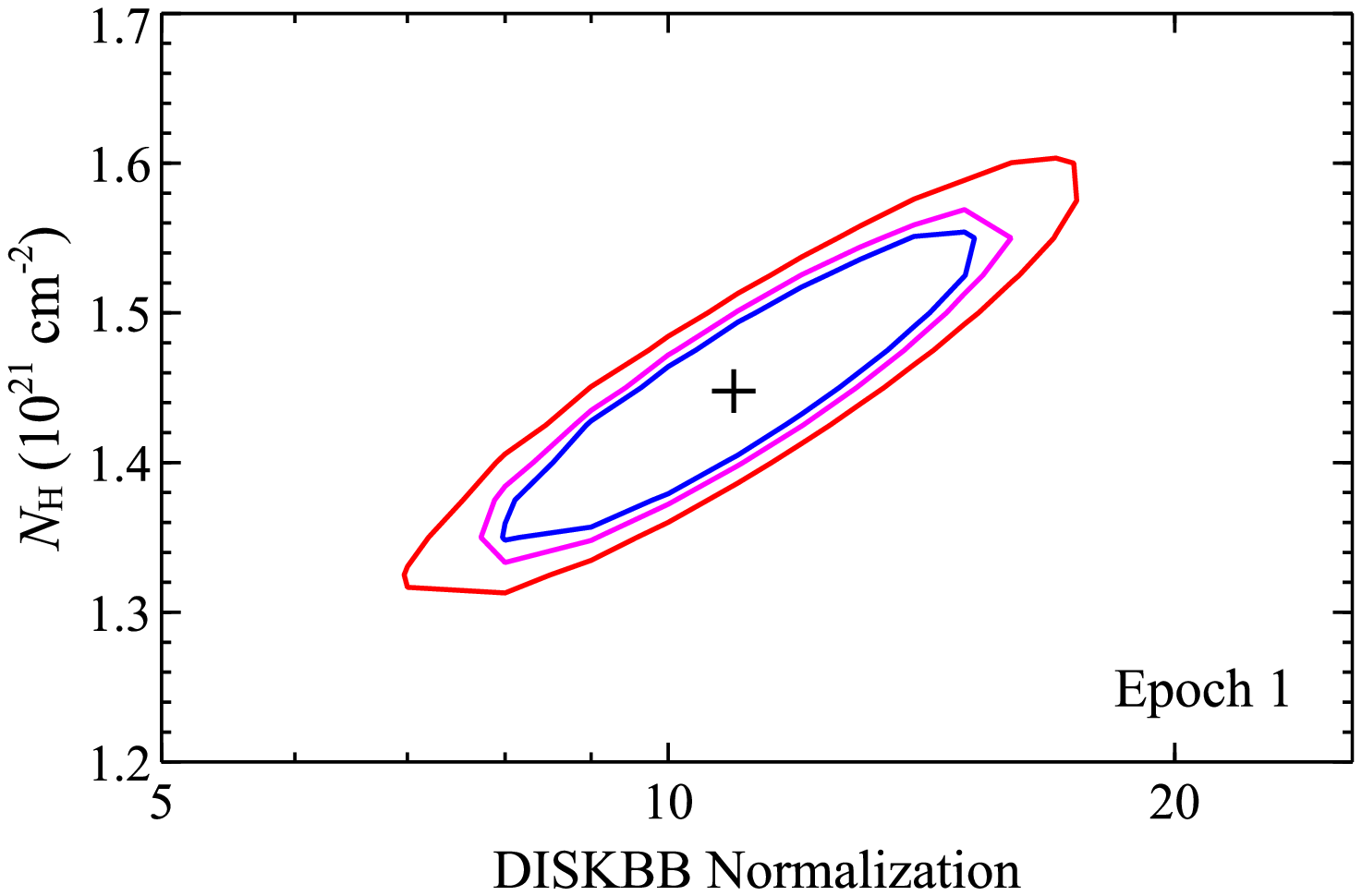}
\caption{\textit{Top panel:} 2D confidence contours for the DISKPBB normalization
from epoch 1 and $f_{\rm{scat}}$ calculated for the scenario with a constant
DISKPBB temperature, showing an example of the general degeneracy between
these parameters that is seen for all epochs. The 90, 95 and 99\% confidence
contours for two parameters of interest are shown in blue, magenta and red,
respectively. \textit{Bottom panel:} An example of the similar degeneracy seen
between the DISKBB normalization from epoch 1 and \nh\ for the scenario in which
the DISKBB temperature is constant.
}
\label{fig_degen}
\end{figure}

Throughout all these fits, we find that in addition to the photon index we can also
link the scattered fraction for the \simpl\ component across all the epochs. With a
constant temperature for the cooler \diskbb\ component, additionally linking
$f_{\rm{scat}}$ gives a fit of \chisq/DoF = 8795/8515, and doing so with a constant
temperature for the hotter \diskpbb\ component gives a fit of \chisq/DoF = 8789/8515
(\ie $\Delta\chi^{2}$ = 3 for another 5 fewer free parameters in both cases). The
best-fit parameters for these two scenarios are given in Table \ref{tab_param}, and
data/model ratio plots for the constant \diskpbb\ temperature scenario are shown for
all six epochs in Figure \ref{fig_specrat}. Although in this scenario the normalizations
for the \diskpbb\ component also appear to be consistent with remaining constant,
this is related to a significant degeneracy between these normalizations and
$f_{\rm{scat}}$ (see Figure \ref{fig_degen}); trying to link this parameter across all
epochs results in a significant degradation in the quality of fit ($\Delta\chi^{2} = 135$
for another 5 fewer free parameters). \simpl\ is a convolution model and
$f_{\rm{scat}}$ indicates the fraction of the input model scattered into the high-energy
continuum, so with our parameterization this acts as a normalization for this
continuum relative to the \diskpbb\ component. As $f_{\rm{scat}}$ is linked across all
epochs, changing this parameter therefore causes each of the \diskpbb\
normalizations to change in tandem (\ie increasing $f_{\rm{scat}}$ causes \textit{all} of
the \diskpbb\ normalizations to increase together). This degeneracy therefore expands
the formal uncertainties on each of the normalizations, but does not remove the need
for the relative variations between epochs seen in the best-fit values.

There is a similar issue between the \diskbb\ normalizations and the neutral absorption
column density in the scenario in which the \diskbb\ temperature is constant. There is
a degeneracy between these parameters for all epochs (again, see Figure
\ref{fig_degen}), and as \nh\ is linked between them changing this parameter again
causes all of the \diskbb\ normalizations to vary in tandem, such that the relative
differences indicated by the best-fit values in Table \ref{tab_param} remain. Both the
thermal components are therefore required to change in at least their normalizations in
both scenarios; the observed spectral variability between these six broadband epochs
cannot be explained by variations in one of these model components only.

Finally, we stress that with the additional data included in this work, the
requirement for the high-energy powerlaw tail is highly significant for any continuum
model that falls away as a Wien tail. If we remove the \simpl\ component and then
allow the temperatures of both of the two thermal components to vary between
epochs, we find a fit of \chisq/DoF = 8875/8512 with clear excesses of emission
seen in the residuals above $\sim$20\,keV. This is worse by $\Delta\chi^{2} > 80$
despite there being three \textit{additional} free parameters in comparison to either
of the fits presented in Table \ref{tab_param}. Furthermore, this is also the case if
we replace the hotter \diskpbb\ with a thermal Comptonization component (\comptt;
\citealt{comptt}), another model that is frequenly considered in the literature.

\begin{figure}
\hspace*{-0.6cm}
\epsscale{1.15}
\plotone{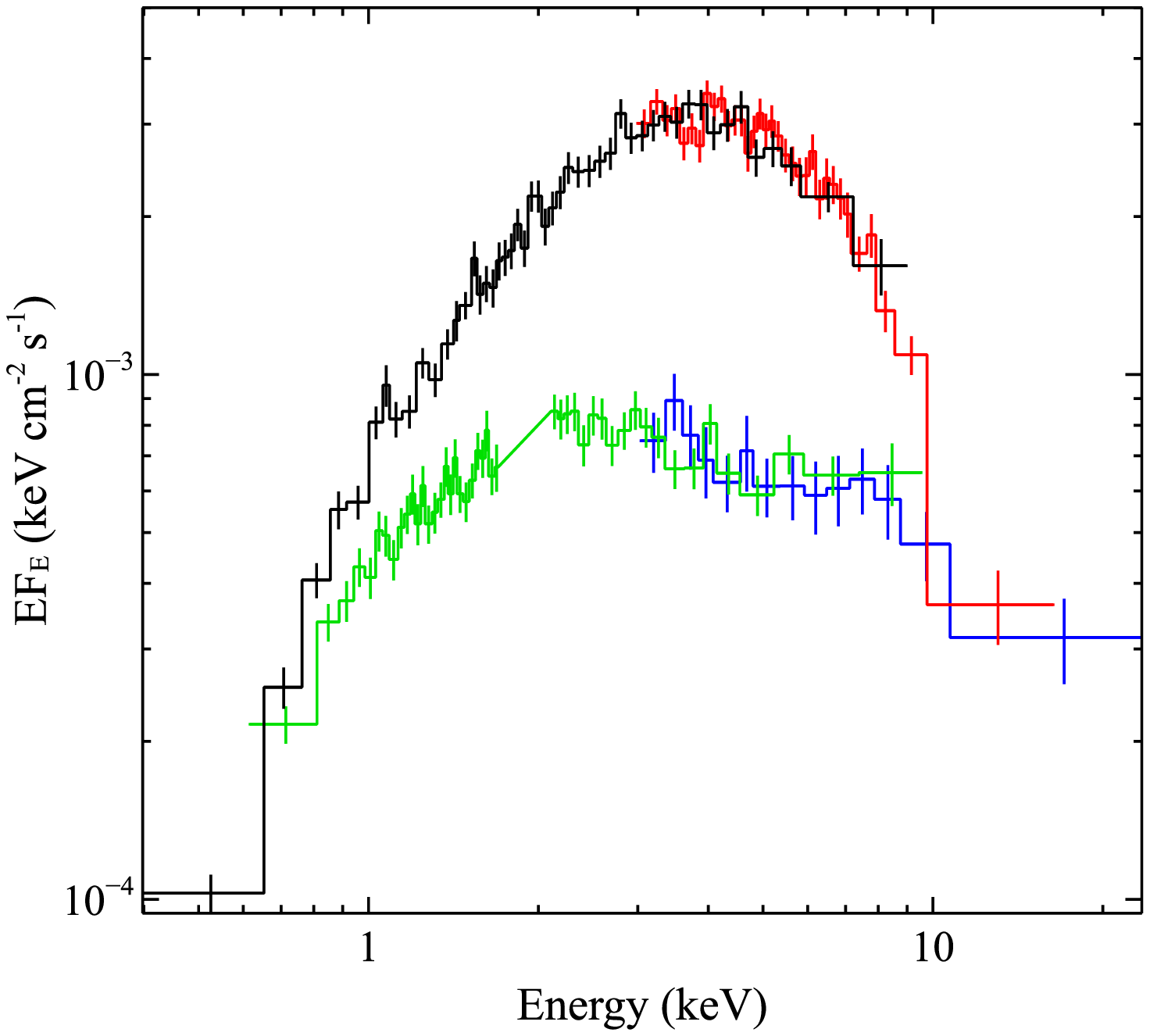}
\caption{The broadband difference spectra computed in Section \ref{sec_diff}. For
clarity, we only show the \epicpn\ and FPMB data for the ``high--med" difference
spectrum (black and red, respectively), and the FI XIS and FPMB data for the
``med--low" difference spectrum (green and blue, respectively). The two difference
spectra are clearly distinct from each other. As with Figure \ref{fig_spec}, all the data
have been unfolded through a model that is constant with energy, and have been
further rebinned for visual purposes.}
\vspace{0.3cm}
\label{fig_diffspec}
\end{figure}

\subsection{Difference Spectroscopy}
\label{sec_diff}

In order to further probe the spectral variability observed, we also compute
`difference spectra' between the three different flux states probed with broadband
observations to date. This involves using a lower flux source spectrum as the
background for a higher flux observation, allowing us to isolate the additional
emission in the latter in a model-independent manner. Given the instrumental
differences, only spectra obtained with the same detectors can be meaningfully
subtracted from one another. While \nustar\ coverage is common among all these
epochs, with respect to the soft X-ray coverage the high and medium flux states
only have \xmm\ coverage in common, while the medium and low flux states only
have the \suzaku\ coverage in common.

In order to compute a broadband difference spectrum between the high and
medium states, we therefore use the \xmm\ and \nustar\ data from epoch 1 as the
background for epoch 2; owing to the different pattern selections for the \epicmos\
data between epochs 1 and 2 to avoid pile-up during epoch 2 (as discussed in
\citealt{Walton14hoIX}), we only consider the \epicpn\ data for \xmm\ here. For the
difference spectrum between the medium and low states, given that the new
observations (epochs 3--6) each have significantly shorter exposures than epochs
1 and 2, we first co-added the \suzaku\ and \nustar\ data from epochs 4 and 5,
and then used these spectra as the background for epoch 1. Given that the
spectra from different epochs are often quite similar, particularly towards the
highest and lowest energies covered here, we substantially increase the bininng
of the data to a minimum of 1000 counts per bin to ensure that the data in the
difference spectra are of sufficient S/N to use \chisq\ minimization. Unfortunately
there is no common soft X-ray instrumental coverage between the high and low
states, so we are not able to compute a broadband difference spectrum in this
case.

The two broadband difference spectra are shown in Figure \ref{fig_diffspec}. Clearly,
the two difference spectra have distinct spectral forms. The ``high--med" (hereafter
HM) difference spectrum is very strongly peaked at $\sim$4\,keV, which is not
surprising given that the epoch 1 and 2 spectra are very similar below $\sim$1\,keV,
and all the spectra are very similar above $\sim$15\,keV. In contrast, the ``med--low"
(hereafter ML) difference spectrum has a much broader profile, but also falls away
at the highest and lowest energies. In order to characterize and further investigate
the differences between them, we proceed to fit these difference spectra with a
variety of simple continuum models. All models also include a neutral absorber
(modeled again with \tbabs) at the redshift of \hoix, but we do not include Galactic
absorption as this should remain constant with time.

We start with the HM difference spectrum. Unsurprisingly, given the strong curvature,
a powerlaw continuum provides an extremely poor fit to the data (\chisq/DoF =
1820/148). Instead, an accretion disk continuum (\diskbb) provides a good fit
(\chisq/DoF = 173/148), capturing this curvature rather well. We also test the \diskpbb\
model, allowing the radial temperature index to be a free parameter, and find a
moderate improvement over the \diskbb\ case (\chisq/DoF = 161/147). However,
in contrast to the \diskpbb\ results for the actual spectra, here we find $p > 0.75$,
implying the emission is slightly more peaked than a standard thin disk continuum. A
run of blackbody temperatures is still required though, as a single temperature
blackbody continuum provides a very poor fit to the data (\chisq/DoF = 706/148).

\begin{figure}
\hspace*{-0.6cm}
\epsscale{1.15}
\plotone{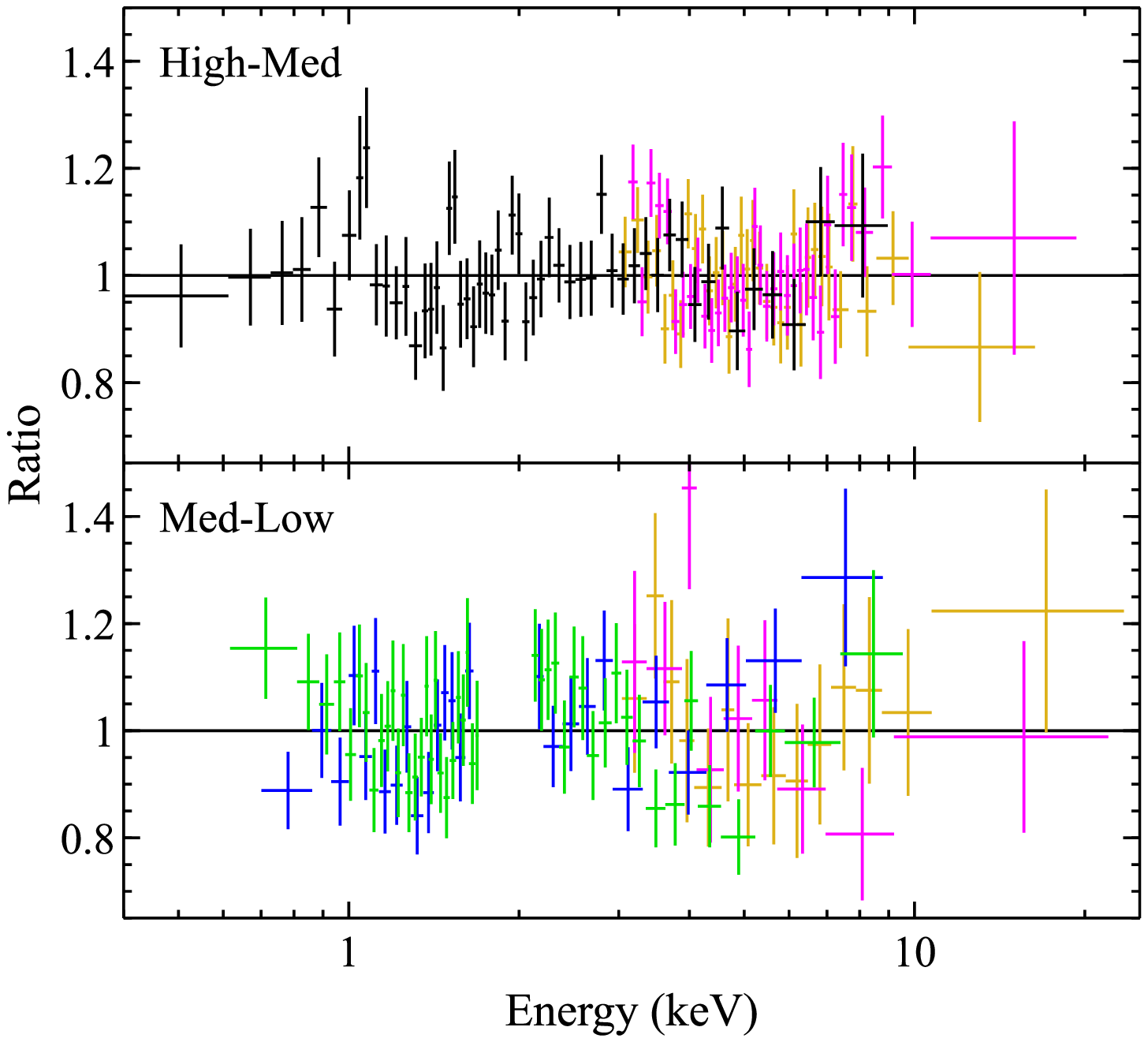}
\caption{Data/model ratios for the two difference spectra considered in Section
\ref{sec_diff} fit with the \diskpbb\ model. The color coding is the same as Figure
\ref{fig_specrat} and again the data have been further rebinned for visual purposes.}
\vspace{0.3cm}
\label{fig_diffrat}
\end{figure}

\begin{table}
  \caption{Results obtained modeling the difference spectra with a simple DISKPBB
  continuum}
\begin{center}
\begin{tabular}{c c c c}
\hline
\hline
\\[-0.15cm]
\multicolumn{2}{c}{Parameter} & High--Med & Med--Low \\
\\[-0.2cm]
\hline
\hline
\\[-0.15cm]
$N_{\rm{H,int}}$ & [$10^{21}$\,cm$^{-2}$] & $2.1^{+0.5}_{-0.4}$ & $3.6 \pm 0.5$ \\
\\[-0.2cm]
$T_{\rm{in}}$ & [keV] & $1.52^{+0.03}_{-0.04}$ & $2.9^{+0.5}_{-0.3}$ \\
\\[-0.2cm]
$p$ & & $0.87^{+0.08}_{-0.06}$ & $0.50 \pm 0.01$ \\
\\[-0.2cm]
Norm & & $0.14^{+0.04}_{-0.03}$ & $(5 \pm 3) \times 10^{-4}$ \\
\\[-0.2cm]
\hline
\\[-0.1cm]
\chisq/DoF & & 161/147 & 490/423 \\
\\[-0.2cm]
\hline
\hline
\\[-0.15cm]
\end{tabular}
\label{tab_diffspec}
\end{center}
\end{table}

For the ML difference spectrum, a powerlaw continuum provides a more acceptable
fit than the HM case (\chisq/DoF = 554/424), but still fails to capture the broad
curvature seen in the data. Here, the \diskbb\ model also provides a poor fit
(\chisq/DoF = 720/424), with the spectral form being significantly less peaked than a
standard thin disk continuum. The \diskpbb\ model again provides a good fit to the
data though (\chisq/DoF = 490/423), with the radial temperature index required to be
significantly flatter than the thin disk case.

We provide the best-fit parameters for the \diskpbb\ model in Table \ref{tab_diffspec}
for direct comparison, and show the data/model ratios for these fits in Figure
\ref{fig_diffrat} as this is the only model to yield good fits for both difference 
spectra.\footnote{Although it visually appears as though there could be an narrow
emission feature in the HM spectrum at $\sim$1~keV, the addition of a Gaussian to
account for this only improves the fit by $\Delta\chi^{2}=7$}. The results are clearly
different for the two difference spectra. We have already highlighted the difference in
the radial temperature index inferred, and the temperatures obtained are also very
different. In addition, the results for the ML difference spectrum also differ significantly
from the best-fit parameters for the \diskbb\ and \diskpbb\ spectral components for
epoch 1 (for both the scenarios considered). While the temperature is similar to the
hotter \diskpbb\ component, $p$ is significantly flatter. This provides further evidence
that the energy and flux-dependent variability shows some complexity, and in the
context of simple dual-thermal models cannot be explained by variations in only one
of the spectral components required to model the observed data.

Finally, we note that the absorption columns obtained for the models that fit the data
well are slightly larger than obtained fitting the actual observed spectra (for the HM
difference spectrum, the column obtained with the \diskbb\ model is slightly higher
again than for the \diskpbb\ model). However, given that the column is consistent
with remaining constant for all epochs when modeling the actual spectra, this does
not likely imply variable absorption (as may be seen, for example, in NGC\,1313 X-1;
\citealt{Middleton15soft}). Instead, this just indicates that the relative variations are
weaker below $\sim$1\,keV than in the $\sim$1--10\,keV range, where the variability
seen in both difference spectra peaks.

\subsection{Fe K Line Search}
\label{sec_FeK}

Lastly, we also re-visit the limits on the presence of any atomic iron absorption or
emission features in light of the ultra-fast outflow ($\sim$0.2--0.25$c$) recently
detected in NGC\,1313 X-1 (\citealt{Pinto16nat, Walton16ufo}). The high-energy
constraints provided by the \nustar\ data, missing in our previous line searches for
\hoix\ (\citealt{Walton12ulxFeK, Walton13hoIXfeK}), mean we can now significantly
improve our sensitivity to any features associated with high-velocity outflows for this
source. Considering the \xmm, \suzaku\ and \nustar\ databases, the majority of the
archival exposure available for \hoix\ covers states similar to epoch 1, and so we
focus on these data in order to maximize the S/N while considering only
observations with similar spectra. Following \cite{Walton16ufo}, we also only
consider \xmm\  observations in which \hoix\ was placed on-axis in order to avoid
increased background emission from copper lines at $\sim$8\,keV seen in the
\epicpn\ detector away from the optical axis (\citealt{XMMblanksky}), which fall in
the energy range of interest. Therefore, in addition to the data from epochs 1, 3
and 6, we also include archival data from a further long \suzaku\ exposure (OBSID
707019010), and two short \xmm\ exposures (ODSIDs 0112521001 and
0112521101). These are reduced in the same manner as the other \xmm\ and
\suzaku\ observations, and then the data are co-added to produce average spectra
for each of the \suzaku\ FI and BI XIS detectors, the \xmm\ \epicpn\ and \epicmos\
detectors, and the \nustar\ FPMA and FPMB modules. The total good exposures
utilized are 570\,ks for each of the operational XIS units, 32\,ks for the \epicpn\
detector, 56\,ks for each of the \epicmos\ detectors, and 315\,ks for each FPM.

For this analysis, we focus on the 3.5--20.0\,keV energy range and model the
continuum local to the iron band with a simple \cutoffpl\ model, as the spectral
curvature in this bandpass is well established (\citealt{Stobbart06, Gladstone09,
Walton13hoIXfeK, Walton14hoIX}). This bandpass is sufficient to accurately model
the continuum local to the iron band; extending to lower energies we find that the
fact that the 0.3--10\,keV energy range requires two components starts to influence
the high-energy continuum fit. We also include the two neutral absorption
components included in our primary spectral analysis (Galactic and intrinsic to
Holmberg\,IX), with the latter fixed to $N_{\rm{H,int}}=1.5\times10^{21}$\,cm$^{-2}$
based on the results above (see also \citealt{Miller13ulx}) since the bandpass
considered is not particularly sensitive to this parameter. This provides an excellent
fit to the 3.5--20.0\,\kev\ emission, with \chisq/DoF = 2287/2269. The photon index
and high-energy cutoff obtained are $\Gamma = 1.07^{+0.04}_{-0.05}$ and
$E_{\rm{cut}} = 7.3 \pm 0.3$\,\kev, and the model normalisation at 1 keV is $(1.22
\pm 0.05) \times 10^{-3}$\,\photpkevpcmsqps.

To search for atomic features, we follow a similar approach to our prior work on ULX
outflows (\eg\ \citealt{Walton13hoIXfeK, Walton16ufo}). The reader is referred to
\cite{Walton12ulxFeK} for more details, but in brief, we slide a narrow ($\sigma =
10$\,eV) Gaussian across the energy range of interest in steps of 40\,eV
(oversampling the \suzaku\ energy resolution by a factor $\sim$4). The Gaussian
normalization is allowed to be either positive or negative (\ie we allow for both
emission and absorption, respectively), and for each line energy we record the
$\Delta\chi^{2}$ improvement in the fit provided by the line, as well as the best-fit
equivalent width ($\mathrm{EW}$) and its 90 and 99\% confidence limits. These
limits are calculated with the {\small EQWIDTH} command in \xspec, using 10,000
simulated  parameter combinations based on the best fit model parameters and their
uncertainties. To be conservative, we vary the line energy between \elow\ and
\ehigh\,\kev, corresponding to a wide range of outflow velocities extending up to
$\sim$0.3$c$ for \fexxvi.

\begin{figure}
\hspace*{-0.5cm}
\epsscale{1.15}
\plotone{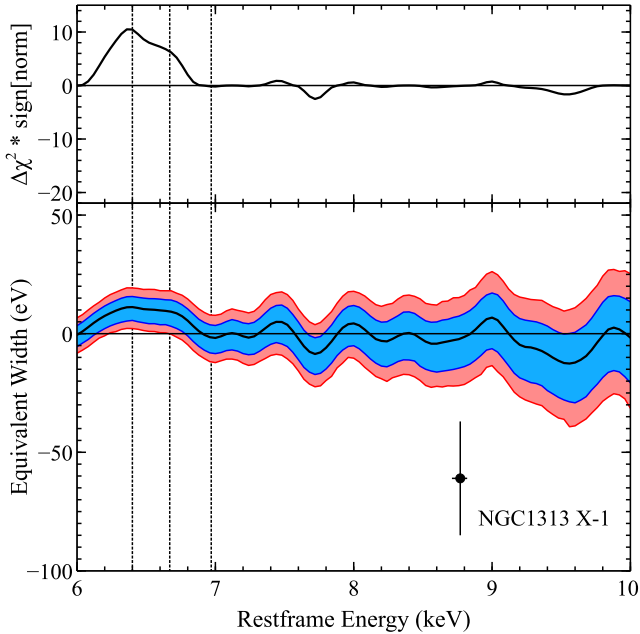}
\caption{Results from our search for narrow atomic iron features in the integrated
spectra of observations similar to the `medium' flux state seen in our broadband
observations (epochs 1, 3 and 6; see Section \ref{sec_FeK}). \textit{Top:} the
$\Delta$\chisq\ improvement provided by a narrow Gaussian line as a function of
rest frame line energy. Positive values of $\Delta$\chisq\ indicate the improvement
is obtained with an emission line, and negative values indicate absorption. There is
a weak indication of excess emission conistent with neutral iron (the rest frame
transitions of neutral, helium-like and hydrogen-like iron are shown with vertical
dashed lines). However, we find no compelling evidence for any narrow absorption
features, even with the additional sensitivity to extreme velocity outflows provided
by the high-energy \nustar\ data. \textit{Bottom panel:} 90\% (\textit{blue}) and 99\%
(\textit{red}) confidence contours for the narrow line equivalent width, indicating for
the majority of the bandpass considered the line strengths any narrow absorption
features present in the data could have and still remain undetected. For context,
we show the high-energy absorption feature detected in NGC\,1313 X-1
(\citealt{Walton16ufo}). Similar absorption along our line of sight to \hoix\ can
confidently be ruled out.
}
\vspace{0.3cm}
\label{fig_search}
\end{figure}

The results are shown in Figure \ref{fig_search}. There is a weak indication of some
iron emission, with a Gaussian emission line providing a moderate improvement
over a range of energies in the immediate iron bandpass. To investigate this further,
we perform additional fits modeling this excess emission first as a single Gaussian
emission line with the energy, width and normalization all free to vary, and second
with two narrow ($\sigma = 10$\,eV) emission lines with energies fixed at the \ka\
transitions for \fei\ and \fexxv\ (6.4 and 6.67\,keV, respectively), and find that both
scenarios fit the data equally well. In the former scenario, the fit is improved by
$\Delta\chi^{2}$/DoF = 18/3 and we find a line energy of $E = 6.44^{+0.14}_{-0.13}$
keV, a line width of $\sigma = 0.24^{+0.24}_{-0.11}$\,keV and an equivalent width of
$\mathrm{EW} = 27 \pm 14$\,eV. In the latter, the fit is improved by by
$\Delta\chi^{2}$/DoF = 16/2, and we find equivalent widths of
$\mathrm{EW}_{\rm{FeI}} = 11 \pm 5$ and $\mathrm{EW}_{\rm{FeXXV}} =
8^{+7}_{-6}$\,eV, respectively.

Based on the well-established relation between the narrow Fe K equivalent width and
the line-of-sight absorption observed in Galactic high-mass X-ray binaries, which is
consistent with expectation for a spherical reprocessing geometry, for an absorption
column of $N_{\rm{H}} = 1.5 \times 10^{21}$\,\pcmsq\ we would expect an equivalent
width of only $\sim$0.5\,eV (\citealt{Torrejon10}). This is significantly smaller than
observed even in the case that the excess emission arises from both \fei\ and \fexxv,
suggesting that the geometry of the reprocessing material cannot be spherical, and
there must be additional reprocessing material out of our line-of-sight in this case.
However, the observed equivalent widths are too small for the reprocessor to be the
accretion disk (\eg\ \citealt{George91}), and the disk temperatures for super-Eddington
accretion onto stellar remnant accretors should be too hot to produce \fei\ emission,
so the line emission likely has some other origin.

More importantly, we find no significant indication of any absorption features, even
after extending our search out to cover extreme outflow velocities. Thanks to the
high-energy coverage from \nustar, the limits on any narrow features present in the
spectrum but undetected by these data remain tight right up to 10\,keV, a significant
improvement on our previous work (\citealt{Walton13hoIXfeK}). The weakest limits, at
$\sim$9.6\,keV, require any undetected absorption features to have $\mathrm{EW} <
40$\,eV at 99\% confidence, but the limits are more typically $\mathrm{EW} < 25$\,eV
across the majority of the bandpass considered. For comparison, we also show the
potential iron absorption feature recently detected in NGC\,1313 X-1 in Figure
\ref{fig_search}. This has an observed energy of $E=8.77^{+0.05}_{-0.06}$\,\kev\ and
an equivalent width of $\mathrm{EW}=-61\pm24$\,eV (\citealt{Walton16ufo}). The
presence of a similar iron absorption component to any outflow along our line of sight
to the hotter continuum emission is ruled out at high ($>$99.9\%) confidence for \hoix.

\section{Discussion and Conclusions}
\label{sec_dis}

We have presented four new broadband (0.6--40.0\,keV) X-ray observations of the
extreme ULX \hoix, obtained with \suzaku\ and \nustar\ in coordination. In addition to
the two initial broadband observations discussed in \citet[][epochs 1 and
2]{Walton14hoIX}, which coordinated \xmm, \suzaku\ and \nustar\ (covering 0.3--40.0
keV), we now have a combined dataset consisting of six epochs on this remarkable
source covering timescales from $\sim$weeks to $\sim$years. The main purpose of
these additional observations was to further investigate the unusual spectral variability
seen between the first two epochs by probing a broader range of fluxes. Of the four
new observations, two (epochs 3 and 6) were found to be broadly similar to the state
seen in epoch 1, and the other two (epochs 4 and 5), which were also similar to each
other) caught \hoix\ in a lower flux state than seen in either epochs 1 or 2. The
spectrum from this lower flux state is again different from the spectra seen from epochs
1 and 2 (see Figure \ref{fig_spec}).

The spectral variability below 10\,keV observed here between these epochs is similar
to that qualitatively described by \cite{Luangtip16}, who present a comprehensive
analysis of the archival X-ray observations of \hoix\ performed by \xmm\ and \swift\
(see also \citealt{Vierdayanti10}). These authors found that at lower fluxes the spectra
appear to show two distinct components peaking at $\sim$1 and $\sim$7\,keV, with
the higher energy component being dominant. Then, as the flux increases, the
curvature between $\sim$1--7\,keV increases significantly, with the spectra strongly
peaking at $\sim$3--4\,keV (for comparison, see their Figure 6). They also found
evidence that the relative level of variability was smaller below $\sim$1\,keV than
above, which is also supported by the observations analysed here.

The key aspect of this work is our ability to extend such analysis into the band above
10\,keV. Remarkably, despite the factor of $\sim$3 variations in flux seen below
10\,keV between the broadband epochs considered here (which probe
$\sim$week--year timescales), the 15--40\,keV band covered by \nustar\ only shows
variations at the $\sim$20\% level. It is well established that the spectra below
10\,keV require two continuum components which resemble multi-color blackbody
emission (\eg\ \citealt{Gladstone09, Walton14hoIX, Luangtip16}). These new
observations robustly confirm that, for any continuum that falls away with a Wien tail,
an additional steep ($\Gamma \sim 3.5$) powerlaw tail is also required to fit the
broadband data. This dominates the 15--40\,keV band in which the relative long-term
stability is seen.\footnote{Note that there is some evidence that the level of variability
on shorter timescales ($\sim$5--70\,ks) may increase above 10\,keV, at least during
epoch 2 (\citealt{Walton14hoIX}). This will be further addressed in future work
(Middleton et al., in preparation).} Evidence for such steep powerlaw tails has now
been seen in several ULXs thanks to observations with \nustar\ (\eg\
\citealt{Walton15hoII, Mukherjee15}).

Building on our initial work on the data from epochs 1 and 2 (\citealt{Walton14hoIX}),
we construct a spectral model consisting of two multicolor blackbody disk components
to model the spectral shape below 10\,keV, as well as a high-energy powerlaw tail.
In order to investigate what drives the observed spectral variability, we apply this model
to the data from each of the six epochs simultaneously. In addition, we also construct
difference spectra between the high and the medium, and the medium and the low flux
states. While our prior work found that the variability between epochs 1 and 2 could be
explained by variations in only one of the two thermal components, with the other
component remaining (and also found that having either of the two as the variable
component provided similarly good fits to the data), when all six epochs are considered
this is no longer the case, and variations in both are required.

Super-Eddington accretion flows are expected to have a large scale-height, owing to
support by the intense radiation pressure (\eg\ \citealt{Shakura73, Abram88,
Poutanen07, Dotan11}), with higher accretion rates resulting in larger scale-heights. 
As such, the inner regions of the accretion flow may more closely resemble a funnel
geometry than a thin disk. \citet[][and references therein]{King16} argue this should
be the case for both neutron star and black hole accretors in the super-Eddington
regime.\footnote{For ULXs in which pulsations are observed, \eg\ M82 X-2
(\citealt{Bachetti14nat}), NGC\,7793 P13 (\citealt{Fuerst16p13, Israel17p13}) and
NGC\,5907 ULX1 (\citealt{Israel16}), some additional complexity must be
present in the inner accretion flow, as this must transition into an accretion column in
order to produce the pulsed emission (see \citealt{Kawashima16}). At the time of
writing, no such pulsations have been observed from \hoix\ by either \xmm\
(\citealt{Doroshenko15}) or \nustar\ (see Appendix \ref{app_pulse}).} 
In addition, strong winds should be launched from the large scale-height regions of
the flow. The observed spectral and variability properties of such flows should have a
strong viewing angle dependence, with the emission from inner regions being
geometrically collimated by the opening angle of the flow, and becoming progressively
more obscured from view by the outer regions when seen at larger inclination angles,
while emission from the outer regions would be more isotropic (\eg\
\citealt{Middleton15}).

Currently one of the most popular interpretations of the different thermal components
required to successfully model ULX spectra below 10\,keV is that the cooler
component represents emission from the outer photosphere (and beyond) of the
funnel formed by the combination of the large scale-height regions of the flow and the
wind launched from these regions, and the hotter component represents emission
from the regions interior to the funnel (\eg\ \citealt{Walton14hoIX, Mukherjee15,
Middleton15, Luangtip16}, although see \citealt{Miller14} for an alternative scenario).
For sources seen at higher inclinations the lower temperature component would be
expected to dominate the observed spectra, while the hotter temperature component 
would dominate for sources seen more face-on. If the winds launched are clumpy, as
predicted by simulations (\eg\ \citealt{Takeuchi13}), the line-of-sight variations in the
wind should result in increased variability, but seen primarily in the higher-energy
component. Among the general ULX population, there is observational evidence that
sources with softer spectra do have stronger short timescale variability, and that this
variability is indeed primarily seen in the hotter of the two components that dominate
below 10\,keV (\citealt{Sutton13uls, Middleton11a, Middleton15}).

In the context of such a model, \cite{Luangtip16} interpret the spectral variability seen
below 10\,keV as arising from changes in the mass accretion rate, with the system
seen close to face-on such that we are looking down the funnel of the flow. Increases
in the accretion rate increase the disk scale-height, causing the opening angle of the
funnel to close. This results in an increase in the degree of geometric beaming
experienced by the hotter emission from within the funnel which, owing to our viewing
angle, causes the flux observed from the hotter component to increase faster than that
from the lower temperature emission. The lack of iron absorption would support the
hypothesis that we are viewing \hoix\ close to face-on during the medium flux state,
particularly in the context of the iron absorption from a massive disk wind seen in
another extreme ULX, NGC\,1313 X-1 (\citealt{Walton16ufo}). Similar iron absorption
along our line of sight can be excluded here (Figure \ref{fig_search}), so if \hoix\ bears
any similarity to NGC\,1313 X-1, we must therefore view it at an angle such that this
wind would not intercept our line of sight to the harder ($>$2\,keV) X-ray emission.
It is possible that the less ionised outer regions of such an outflow could explain the
iron emission observed.

In the scenario in which a super-Eddington accretion disk dominates the emission
below 10\,keV, the most natural explanation for the steep high-energy powerlaw
component required by such models is that it arises through Compton up-scattering of
the blackbody emission from the accretion flow in a corona of hot electrons, similar to
the high-energy powerlaw tails seen in both Galactic X-ray binaries and active
galactic nuclei. A steep spectrum would likely be expected for such emission at
high/super-Eddington accretion rates (\eg\ \citealt{Risaliti09LxGam, Brightman13,
Brightman16LxGam}). Alternatively the high-energy tail could be related to Compton
scattering in the accretion column should the central compact object be a neutron star,
or even potentially bulk motion Comptonization in a super-Eddington wind launched
from the accretion flow; evidence for such winds has now been observed from a
handful of ULX systems through both emission and absorption features
(\citealt{Pinto16nat, Pinto17, Walton16ufo}).

The relative lack of long-term variability seen from these energies in comparison to that
seen from the emission below $\sim$10\,keV may therefore be problematic for the
scenario proposed by \cite{Luangtip16}. X-ray coronae in sub-Eddington black hole
systems are generally understood to be very compact ($\lesssim$20\,\rg, where \rg\ =
$GM/c^{2}$ is the gravitational radius; \eg\ \citealt{Reis13corona}, and references
therein). Although it is possible the corona could be different in the super-Eddington 
regime, should there be any similarity with the sub-Eddington regime this emission
would be expected to arise from within the funnel formed by the accretion flow/disk
wind for a black hole accretor. This would naturally also be the case should this
emission be associated with a neutron star accretion column. In both cases, the
high-energy powerlaw emission would then be subject to similar geometrical beaming
as the blackbody emission from within this region. However, the relative levels of
long-term variability seen above and below 10\,keV would strongly suggest that this is
not the case. The high-energy stability may also be problematic for the bulk motion
Comptonization possibility. As the mass loss in the wind should naturally increase with
increasing accretion rate through the disk, and thus the observed flux at lower
energies, we would again also expect the high-energy emission to respond accordingly.

\cite{Middleton15} also discuss the possibility of precession of the accretion flow
introducing spectral variability. When the system is viewed close to the opening angle
of the flow, precession can change the level of self-obscuration of the inner regions of
the flow by the outer regions. However, in this case we are likely viewing the system
close to face-on even during the medium-flux epochs, making it more difficult for
precession to have a significant effect in this manner. Furthermore, the majority of the
variability is seen at intermediate energies/temperatures in terms of the broadband
spectra obtained ($\sim$1--10\,keV); if the inner regions of the flow were being
blocked from our view at times, and exposed at other times, one would likely again
expect to see the highest energies respond similarly. Indeed, in the case of NGC\,5907
ULX1, where 78\,d super-orbital\footnote{The orbital period is now known to be
$\sim$5\,d (\citealt{Israel16}).} variations that may well be related to precession of the
accretion flow are observed (\citealt{Walton16period}), there is evidence that
high-energy flux does vary along with the lower-energy emission
(\citealt{Fuerst17ngc5907}; although observations are still sparse).

We suggest the observed broadband data could potentially be explained with a small
but perhaps important adjustment to the model discussed by \cite{Luangtip16}. Instead
of simply closing the opening angle of an otherwise static inner funnel, consider the
case in which the main effect of an increasing mass accretion rate (potentially in
addition to increasing the scale-height of the outer flow) is that the effective radius
within which strong geometric beaming occurs moves outwards. This may not be
unreasonable, as the characteristic radius from which the wind (which helps to form
the funnel) is launched is expected to increase with accretion rate in most
super-Eddington models (\eg\ \citealt{Shakura73, Poutanen07}). In such a scenario,
as the accretion rate increases, the beamed region would expand to include gradually
lower temperature regions of the flow, potentially without significantly influencing the
degree to which the innermost regions are beamed. Such an evolution in the flow
could potentially explain both the relative long-timescale stability of the highest
energy/temperature emission, which is always beamed, and the strong variability at
intermediate energies/temperatures as the emission from these regions of the disk
becomes more beamed. It would also be broadly consistent with the fact that, when fit
with simple thermal models, the temperature obtained for the ML difference spectrum
is higher than for the HM difference spectrum.

Furthermore, this picture also predicts that as the accretion rate increases, the regions
of the accretion flow exterior to the inner funnel should move progressively to larger
radii. In the models applied to the data here, these regions would be associated with
the lower-temperature \diskbb\ component, and the behaviour of this emission is best
judged by the scenario in which the \diskbb\ temperature is constant. The \diskbb\
normalization is proportional to $R_{\rm{in}}^{2}/f_{\rm{col}}^{4}$, where $R_{\rm{in}}$
is the inner radius of the disk and $f_{\rm{col}} = T_{\rm{col}}/T_{\rm{eff}}$ is the color
correction factor relating the observed `color' temperature to the effective temperature
at the midplane of the disk, providing a simple empirical correction to accounting for
the complex physics in the disk atmosphere. Assuming a constant $f_{\rm{col}}$,
which should be reasonable for the regions exterior to the funnel where the disk
structure is expected to be more standard, the relative evolution of the normalization
does indeed suggest that this component arises from larger radii at higher fluxes (see
Table \ref{tab_param}). We also note that the absolute radii inferred from the \diskbb\
component, imply that, even if the central accretor is a neutron star, this emission
component cannot arise from the surface of the neutron star itself, further supporting
its identification as the outer accretion flow. Even assuming no color correction and a
viewing angle of $\cos\theta = 1$, which would give the smallest radius, we find a
typical inner radius of $\sim$10$^{6}$\,km, orders of magnitude larger than neutron
star radii.

In this scenario, the total amount of beamed emission would still increase with
increasing mass accretion rate, as broadly expected (\eg\ \citealt{King09}), as a
progressively larger fraction of the observed emission arises from within the region
in which geometric collimation is important. However, keeping the degree to which
the emission from the very innermost regions is collimated relatively constant would
likely require that the opening angle of the funnel does not change significantly
between these observations. Obviously if the flow is to transition from a thin disk at
sub-Eddington luminosities to a thicker, funnel-like disk at super-Eddington
luminosities then the opening angle of the funnel must close with increasing
accretion rate over some range of accretion rate. However, an interesting possibility
is that this evolution saturates as the accretion rate continues to increase. While the
increased radiation pressure within the disk itself acts to increase its scale height,
once this scale-height is significant, radiation pressure from the opposite face of the
funnel must also act against any further closing of its opening angle, which could
potentially help produce this effect. Furthermore, \cite{Lasota16} argue that at the
highest accretion rates the disk may become fully advection-dominated, and that in
this regime the scale-height of the disk can no longer increase (or equivalently, the
opening angle of the inner funnel can no longer close) with increasing accretion rate,
as basically all the energy is advected over the horizon and is not able to further
inflate the disk. A roughly constant opening angle for the inner disk, as may be
required to explain the broadband \hoix\ observations presented here, could
therefore be a plausible physical scenario.

\section*{ACKNOWLEDGEMENTS}

The authors would like to thank the anonymous reviewer, who provided useful
suggestions for improving the final manuscript, and also Tim Roberts for useful
discussion. DJW and MJM acknowledge support from STFC through Ernest
Rutherford fellowships, ACF acknowledges support from ERC Advanced Grant
340442, and DB acknowledges financial support from the French Space Agency
(CNES). This research has made use of data obtained with \nustar, a project led by
Caltech, funded by NASA and managed by NASA/JPL, and has utilized the
\nustardas\ software  package, jointly developed by the ASDC (Italy) and Caltech
(USA). This research has also made use of data obtained with \xmm, an ESA
science mission with instruments and contributions directly funded by ESA Member
States, and with \suzaku, a collaborative mission between the space agencies of
Japan (JAXA) and the USA (NASA).

{\it Facilites:} \facility{NuSTAR}, \facility{XMM}, \facility{Suzaku}

\appendix

\section{A. \textit{NuSTAR} Pulsation Search}
\label{app_pulse}

We calculated power spectral densities (PSDs) for all ObsIDs based on 3--40\,keV
light curves with 0.05\,s resolution. All times were transferred to the solar barycenter
using the DE200 solar ephemeris. We searched for fast, coherent pulsations
between 6\,mHz and 10\,Hz in all ObsIDs separately, averaging over at least 37
PSDs per epoch, similar to the approach taken in \cite{Fuerst16p13}. No significant
excess over the Poisson noise was found in any ObsID.

\bibliographystyle{/Users/dwalton/papers/mnras}

\bibliography{/Users/dwalton/papers/references}

\label{lastpage}

\end{document}